\newcommand\apjcls{1}
\newcommand\aastexcls{2}
\newcommand\othercls{3}

\newcommand\papercls{\aastexcls}
\documentclass[tighten, times, preprint2]{aastex6}  

\if\papercls \apjcls
\usepackage{apjfonts}
\else\if\papercls \othercls
\usepackage{epsfig}
\fi\fi
\usepackage{ifthen}
\usepackage{natbib}
\usepackage{amssymb, amsmath}
\usepackage{appendix}
\usepackage{etoolbox}
\usepackage[T1]{fontenc}
\usepackage{paralist}

\if\papercls \apjcls
\newcommand\aas{\ref@jnl{AAS Meeting Abstracts}}
\newcommand\dps{\ref@jnl{AAS/DPS Meeting Abstracts}}
\newcommand\maps{\ref@jnl{MAPS}}
\else\if\papercls \othercls
\usepackage{astjnlabbrev-jh}
\fi\fi

\if\papercls \aastexcls
\hypersetup{citecolor=blue, 
            linkcolor=blue, 
            menucolor=blue, 
            urlcolor=blue}  
\else
\usepackage[
bookmarks=true,           
bookmarksnumbered=true,   
colorlinks=true,          
citecolor=blue,           
linkcolor=blue,           
menucolor=blue,           
urlcolor=blue,            
linkbordercolor={0 0 1},  
pdfborder={0 0 1},
frenchlinks=true]{hyperref}
\fi

\if\papercls \othercls

\else

\fi

\providecommand{\adsurl}[1]{\href{#1}{ADS}}

\makeatletter
\patchcmd{\NAT@citex}
  {\@citea\NAT@hyper@{%
     \NAT@nmfmt{\NAT@nm}%
     \hyper@natlinkbreak{\NAT@aysep\NAT@spacechar}{\@citeb\@extra@b@citeb}%
     \NAT@date}}
  {\@citea\NAT@nmfmt{\NAT@nm}%
   \NAT@aysep\NAT@spacechar\NAT@hyper@{\NAT@date}}{}{}

\patchcmd{\NAT@citex}
  {\@citea\NAT@hyper@{%
     \NAT@nmfmt{\NAT@nm}%
     \hyper@natlinkbreak{\NAT@spacechar\NAT@@open\if*#1*\else#1\NAT@spacechar\fi}%
       {\@citeb\@extra@b@citeb}%
     \NAT@date}}
  {\@citea\NAT@nmfmt{\NAT@nm}%
   \NAT@spacechar\NAT@@open\if*#1*\else#1\NAT@spacechar\fi\NAT@hyper@{\NAT@date}}
  {}{}
\makeatother

\makeatletter
\DeclareRobustCommand{\lowcase}[1]{\@lowcase#1\@nil}
\def\@lowcase#1\@nil{\if\relax#1\relax\else\MakeLowercase{#1}\fi}
\makeatother

\DeclareSymbolFont{UPM}{U}{eur}{m}{n}
\DeclareMathSymbol{\umu}{0}{UPM}{"16}
\let\oldumu=\umu
\renewcommand\umu{\ifmmode\oldumu\else\math{\oldumu}\fi}

\if\papercls \othercls

\else

\fi

\let\oldsim=\sim
\renewcommand\sim{\ifmmode\oldsim\else\math{\oldsim}\fi}
\let\oldpm=\pm
\renewcommand\pm{\ifmmode\oldpm\else\math{\oldpm}\fi}
\newcommand\by{\ifmmode\times\else\math{\times}\fi}

\newcommand\tablebox[1]{\begin{tabular}[t]{@{}l@{}}#1\end{tabular}}
\newbox{\wdbox}
\renewcommand\c{\setbox\wdbox=\hbox{,}\hspace{\wd\wdbox}}
\renewcommand\i{\setbox\wdbox=\hbox{i}\hspace{\wd\wdbox}}

\newcount\timect
\newcount\hourct
\newcount\minct
\newcommand\now{\timect=\time \divide\timect by 60
         \hourct=\timect \multiply\hourct by 60
         \minct=\time \advance\minct by -\hourct
         \number\timect:\ifnum \minct < 10 0\fi\number\minct}

\catcode`@=11

\newcommand\comment[1]{}

\newcommand\commenton{\catcode`\%=14}

\renewcommand\math[1]{$#1$}
\newcommand\mathshifton{\catcode`\$=3}

\let\atab=&
\newcommand\atabon{\catcode`\&=4}

\let\oldmsp=\sp
\let\oldmsb=\sb
\def\sp#1{\ifmmode
           \oldmsp{#1}%
         \else\strut\raise.85ex\hbox{\scriptsize #1}\fi}
\def\sb#1{\ifmmode
           \oldmsb{#1}%
         \else\strut\raise-.54ex\hbox{\scriptsize #1}\fi}
\newbox\@sp
\newbox\@sb
\def\sbp#1#2{\ifmmode%
           \oldmsb{#1}\oldmsp{#2}%
         \else
           \setbox\@sb=\hbox{\sb{#1}}%
           \setbox\@sp=\hbox{\sp{#2}}%
           \rlap{\copy\@sb}\copy\@sp
           \ifdim \wd\@sb >\wd\@sp
             \hskip -\wd\@sp \hskip \wd\@sb
           \fi
        \fi}
\def\msp#1{\ifmmode
           \oldmsp{#1}
         \else \math{\oldmsp{#1}}\fi}
\def\msb#1{\ifmmode
           \oldmsb{#1}
         \else \math{\oldmsb{#1}}\fi}

\def\supon{\catcode`\^=7}

\def\subon{\catcode`\_=8}

\def\supsubon{\supon \subon}

\newcommand\actcharon{\catcode`\~=13}

\newcommand\paramon{\catcode`\#=6}

\comment{And now to turn us totally on and off...}

\newcommand\reservedcharson{ \commenton  \mathshifton  \atabon  \supsubon 
                             \actcharon  \paramon}

\catcode`@=12
\reservedcharson

\if\papercls \apjcls

\else

\fi

\if\papercls \othercls
\else
  \newcommand\inpress{n}
  \if\inpress y
    \received{\today}
    \revised{}
    \accepted{}
    \if\papercls \apjcls
    \slugcomment{}
    \fi
  \else
  \slugcomment{\tablebox{In preparation for {\em ApJ}. DRAFT of {\today}.}}
  \fi
\fi

\newcommand\chisq{\ifmmode{\chi\sp{2}}\else\math{\chi\sp{2}}\fi}
\newcommand\redchisq{\ifmmode{ \chi\sp{2}\sb{\rm red}}
                    \else\math{\chi\sp{2}\sb{\rm red}}\fi}
\newcommand\Teq{\ifmmode{T\sb{\rm eq}}\else$T$\sb{eq}\fi}
\newcommand\mjup{\ifmmode{M\sb{\rm Jup}}\else$M$\sb{Jup}\fi}
\newcommand\rjup{\ifmmode{R\sb{\rm Jup}}\else$R$\sb{Jup}\fi}
\newcommand\msun{\ifmmode{M\sb{\odot}}\else$M\sb{\odot}$\fi}
\newcommand\rsun{\ifmmode{R\sb{\odot}}\else$R\sb{\odot}$\fi}
\newcommand\mearth{\ifmmode{M\sb{\oplus}}\else$M\sb{\oplus}$\fi}
\newcommand\rearth{\ifmmode{R\sb{\oplus}}\else$R\sb{\oplus}$\fi}

\begin{document}

\shorttitle{Boxy Orbital Structures}

\title{Boxy Orbital Structures in Rotating Bar Models}

\author{
L. Chaves-Velasquez\altaffilmark{1}, P.A. Patsis\altaffilmark{2}, I.
Puerari\altaffilmark{1}, Ch. Skokos\altaffilmark{3}, and T. 
Manos\altaffilmark{4,5,6} 
}

\affil{
\sp{1} Instituto Nacional de Astrof\'isica, 
\'Optica y Electr\'onica, Calle Luis Enrique Erro 1, 72840 Santa Mar\'{\i}a
Tonantzintla, Puebla, Mexico\\ 
\sp{2} Research Center for Astronomy, Academy of Athens, Soranou
Efessiou 4, GR-115 27 Athens, Greece\\
\sp{3} Department of Mathematics and Applied Mathematics, University
 of Cape Town, Rondebosch, 7701, South Africa\\ 
\sp{4}CAMTP - Center for Applied Mathematics and Theoretical Physics, University
of Maribor, Krekova 2, SI-2000 Maribor, Slovenia\\ 
\sp{5}School of Applied Sciences, University of Nova Gorica, Vipavska 11c,
SI-5270 Ajdov\v{s}\v{c}ina, Slovenia\\
\sp{6} Institute of Neuroscience and Medicine, Brain and Behaviour (INM-7),
Research Centre J\"{u}lich, J\"{u}lich, Germany\\
}

\email{leonardochaves83@gmail.com}

\begin{abstract}
We investigate regular and chaotic two-dimensional (2D) and three-dimensional 
(3D) orbits of stars in models of a galactic potential consisting in a disk, a 
halo and a bar, to find the origin of boxy components, which are part of the 
bar 
or (almost) the bar itself. Our models originate in snapshots of an 
$N$-body simulation, which develops a strong bar. We consider three snapshots 
of 
the simulation and for the orbital study we treat each snapshot independently, 
as an autonomous Hamiltonian system. The calculated corotation-to-bar-length 
ratios indicate that in all three cases the bar rotates slowly, while the 
orientation of the orbits of the main family of periodic orbits changes along 
its characteristic. We characterize the orbits as regular, sticky, or chaotic 
after integrating them for a 10 Gyr period by using the GALI$_2$ index. 
Boxiness 
in the equatorial plane is associated either with quasi-periodic orbits in the 
outer parts of stability islands, or with sticky orbits around them, which can 
be found in a large range of energies. We indicate the location of such orbits 
in diagrams, which include the characteristic of the main family. They are 
always found about the transition region from order to chaos. By perturbing 
such 
orbits in the vertical direction we find a class of 3D non-periodic orbits, 
which have boxy projections both in their face-on and side-on views.
\end{abstract}

\keywords{Galaxies: kinematics and dynamics -- Galaxies: structure -- chaos}

\section{INTRODUCTION}
\label{introduction}
Strong bars are observed in optical images of almost half of all nearby disk
galaxies \citep[see e.g.][]{2008ApJ...675.1194B, 2007ApJ...659.1176M,
2007AJ....133.2846R}. This percentage increases nearly to $70$~\% when
near-infrared images are considered \citep{2000ApJ...529...93K,
2007ApJ...657..790M, 2000AJ....119..536E}. Bars are characterized by three
parameters: length, strength, and pattern speed. This last parameter is defined
as the rotational frequency of the bar and determines to a large extent the
dynamics of a barred galaxy.  Bars are classified as fast or slow by means of
the ratio $R=R_{CR}/a_{b}$, where $R_{CR}$ is the corotation radius, and $a_{b}$
is the length of the semi-major axis of the bar. The orbital theory shows that
bars cannot extend beyond corotation \citep{1980A.A....81..198C}. In the case of
fast rotators we have $1.0<R<1.4$, while for slow rotators $R>1.4$
\citep{1992MNRAS.259..345A,2000ApJ...543..704D}. By definition, in a slow
rotator, corotation is located far from the end of the bar.

Structures in barred galaxies have to be supported by stellar orbits
\citep{2002ocda.book.....C, 2008gady.book.....B}. It is now known that not only
regular, but chaotic, sticky orbits as well can be used for building the
bars \citep{1994LNP...430..264W,1996A.A...309..381K, 1997ApJ...483..731P,
1999CeMDA..73..149W, 2005CeMDA..91..173M, 2009MNRAS.394.1605H,
2010ASPC..424..377H, 2010MNRAS.408...22P, 2013MNRAS.436.1201C,
2014MNRAS.445.3546P, 2015MNRAS.448.3081T}. Sticky orbits are chaotic orbits
which wander for relatively long times close to the outer borders of stability
islands before eventually entering a well defined chaotic region in the system's
phase space. In some other cases there is also stickiness near unstable
asymptotic curves in the chaotic sea, which is called ``stickiness in chaos''
\citep{2008IJBC...18.2929C}. In both cases, sticky orbits mimic the
behaviour of quasi-periodic orbits in the configuration space during the time
they remain confined in a region of phase space. However, ultimately, during
their time evolution they will exhibit a change in their orbital morphology as
they will at a certain time change their behavior from quasi-regular to
completely chaotic.

Special features and deviations from the standard orbital dynamics 
\citep{1989A.ARv...1..261C} have been encountered in several cases. For 
example, 
in \citet{2015MNRAS.448.3081T} the orbital stellar dynamics of a 
two-dimensional 
(2D), slowly rotating, barred-spiral model has been investigated. In this case, 
orbital families have been presented that support in the galactic plane an 
inner 
ring and an X feature embedded in the bar. However, the dynamics associated 
with 
this model is different from that of a typical bar ending close to corotation. 
The ring was a result of a folding of the characteristic (``S'' shape), along 
which the orientation of the elliptical orbits of the main family, as well as 
their stability vary (bistable bifurcation). On the other hand the observed 
boxiness and the X feature reflected the presence of sticky orbits at energy 
levels corresponding to the middle of the barred-spiral potential. Folding of 
the characteristic curve of the main planar family has been found earlier in 
the 
work of \citet{2002MNRAS.333..861S} in the case of a 3D bar rotating again 
slowly. A question that arises is how common this feature is in the backbone 
families of real bars and what are the implications for the observed 
morphologies.

The aim of this work is to study the underlying dynamics in three analytic
models that have a common origin, being derived from snapshots of an
$N$-body simulation reported in \citet{2010MNRAS.406.2386M}. We want to 
examine the degree of chaoticity of the bar-supporting orbits. In that 
simulation the interaction between the dark matter halo and the disk 
develops a bar, which evolves in time. We consider for our study three
snapshots at times 4.2, 7 and 11.2~Gyr. Each snapshot is modeled
by a frozen potential and so we treat each one of them as a time independent
model, i.e. we use in our work the formalism for autonomous Hamiltonian systems.
The bar is modeled with a Ferrers potential \citep{Fer1877}, the disk is a
Miyamoto-Nagai disk \citep{1975PASJ...27..533M} and the dark matter halo is
these snapshots have been taken from \citet{2014MNRAS.438.2201M} (hereafter MM).
Throughout the text, by referring to a ``snapshot'' we will refer to the
corresponding model in MM.

In particular we want to examine the relation between morphological features of
the bars and the degree of chaos of the orbits that support these features. Such
features include a possible inner and/or outer boxiness of the bar and
the formation of rings. In the 3D barred models in \citet{2014MNRAS.445.3525P,
2014MNRAS.445.3546P} it has been suggested that \textit{inner} boxy features can
be built by means of quasi-periodic orbits at the edges of the stability islands
of the x1 family, as well as with sticky orbits just beyond the last invariant
torus around the stable x1 periodic orbit. It has been also proposed that such
orbits support boxiness both in face-on, as well as in edge-on projections
at the central region of the bar (about within half the way to the end
of the bar). A similar dynamical phenomenon was leading to the boxy features on
the galactic plane in the bars of 2D barred-spiral models in
\citet{2015MNRAS.448.3081T}. 

In the present study we want to investigate what kind of orbits support double
boxy morphologies in the successive snapshots, and how they evolve in time, i.e.
from the model of the earlier snapshot to the model for the final one.
We want to examine whether this dynamical mechanism is associated 
with orbits just beyond the vertical 2:1 resonance region, or can be applied in
a large energy range in which we can find bar-supporting orbits. For this
purpose we do not investigate in detail the structure of phase space in a large
number of energies, but we investigate the systems' \textit{global} dynamics
using chaos indicators.

Many techniques have been developed over the years for determining the regular
or chaotic nature of orbits of dynamical systems. Review presentations of some
of the most commonly used methods can be found in \citet{2016LNP...915E...1S}.
Among these chaos indicators the Smaller Alignment Index (SALI) method
\citep{2001JPhA...3410029S,2003PThPS.150..439S,2004JPhA...37.6269S} and its
extension, the Generalized Alignment Index (GALI) technique
\citep{2007PhyD..231...30S, 2008EPJST.165....5S, 2012IJBC...2250218M} proved to
be quite efficient in revealing the chaotic nature of orbits of Hamiltonian
systems in a fast and accurate way. The computation of these indices is based on
the time evolution of more than one deviations from the studied orbits. The
SALI/GALI methods have already been successfully applied to dynamical studies of
astronomical problems \citep[see
e.g.][]{2004CeMDA..90..127S, 2007ApJ...666..165C, 2007CeMDA..99..129S,
2007MNRAS.381..757V, 2008NPCS...11..171M, 2008ApJ...675..802V,
2009CeMDA.104..205B, 2009MNRAS.394.1605H, 2011MNRAS.415..629M,
2013JPhA...46y4017M, 2014MNRAS.438.2871C, 2016MNRAS.458.3578M}.
The reader is
referred to \citet{2016Springer} for a recent review of the theory and 
applications of the SALI/GALI chaos indicators. 

In order to study the degree of chaoticity of the orbits of our models we use 
the GALI$_2$ index, whose time evolution reveals quite efficiently the regular, 
sticky or chaotic nature of the studied orbit. It can also tell the time 
interval within which a sticky orbit behaves as a regular one, being able this 
way to support a given morphological structure. For these reasons the use of 
GALI$_2$ is an essential tool for the needs of our investigation. We also note 
here that the GALI$_2$ index is closely related to the SALI method \citep[see 
for example Appendix B of][]{2007PhyD..231...30S}.

The paper is organized as follows: In section \ref{sec:Nbody_ss} we explain 
the gravitational potentials that model the components of the $N$-body snapshots.
In section \ref{sec:Autonomous Hamiltonian system and the GALI$_2$ index} 
we present the numerical methods used in our 
study. In particular we introduce the Hamiltonian of the system. The GALI$_2$ index is 
introduced as well. In section \ref{degchaos} we present the results of our study. We describe the 
characteristic curves of the main planar family of periodic orbits in the models 
we study and we label the initial conditions of the integrated non-periodic 
orbits, according to the degree of their chaoticity. Finally in section 
\ref{sec:conclusions} we summarize our findings and we present and discuss our 
conclusions.

\vspace{1cm}
\section[]{Modelling the $N$-body snapshots}
\label{sec:Nbody_ss}
In our study we follow closely the approach of the work of
MM. The models used for approximating the morphologies encountered in the
studied snapshots of the \hbox{$N$-body} simulation consist of a bar embedded 
in 
an
axisymmetric disk and halo environment. The bar is represented by a Ferrers 
model (Ferrers 1877), the disk is a typical Miyamoto-Nagai model 
\citep{1975PASJ...27..533M} and the spherical dark mater halo surrounding the 
disk is represented by a
Dehnen potential \citep{1993MNRAS.265..250D}. The mathematical formulae for 
these potentials can be found in MM.


The structural and dynamical parameters of the bar, disk and halo of the models
are adopted from the models in MM and are summarized in Table~\ref{tabparams}.
In this table we include also an earlier snapshot presented in MM, at $t=1.4$
Gyr, which however has not developed yet a strong bar. We keep it in the table,
but we will not
present any orbital analysis for its small bar. Thus, the models in
Table~\ref{tabparams} correspond to four snapshots taken at times $t=1.4$ Gyr,
$t=4.2$ Gyr, $t=7.0$ Gyr, and $t=11.2$ Gyr. We name them snapshot ``1'', ``2'',
``3'' and ``4'' respectively. 


The scaling of units we used in our calculations, which corresponds also to the
numbers that appear in the axes of the figures in this work, are: 1~kpc 
(length),
$2\times10^{11} M_\odot$ (mass), 1 kpc$^2$/Myr$^2$ (energy), while G = 1.


\begin{table*}
\centering \footnotesize \caption{The parameters of the models fitting the
snapshots of the $N$-body simulation of \citet{2014MNRAS.438.2201M}.
Successively we
give the number of the snapshot (s/s), the time of the snapshot, the semi-axes
of the Ferrers bar $(a,b,c)$, the pattern speed of the bar $\Omega_b$, the mass
of the bar $M_B$, the scale lengths of the Miyamoto disk $A$ and $B$, the disk
mass
$M_D$, the scale radius of the halo $a_H$, the dimensionless parameter $\gamma$ 
in the Dehnen halo potential and the mass of the halo $M_H$ (units as in the MM 
paper).}
\label{tabparams}
\begin{tabular}{c c c c c c c c c c c c c c c c}
\hline
 & & & & Bar & & & & & Disk & & & & Halo & \\
\hline
  \textbf{s/s} & time  & & $a$ & $b$ & $c$ & $\Omega_b$ & $M_B$ & &$A$ & $B$ &
$M_D$ & & $a_H$
& $\gamma$ & $M_H$ \\
 & (Gyr)  & & (kpc) & (kpc) & (kpc) & (km s$^{-1}$ kpc$^{-1}$) & ($10^{10}
M_{\odot}$) & & (kpc) & (kpc) & ($10^{10} M_{\odot}$) & & (kpc) & &  ($10^{10}
M_{\odot}$)  \\
\hline
\textbf{1}& ~1.4 & & 2.24 & 0.71 & 0.44 & 52 & 1.04 & & 1.92 & 0.22 & 3.96 & &
3.90 & 0.23 &
25 \\
\textbf{2}& ~4.2 & & 5.40 & 1.76 & 1.13 & 24 & 2.36 & & 0.95 & 0.53 & 2.64 & &
5.21 & 0.71 &
25 \\
\textbf{3}& ~7.0 & & 7.15 & 2.38 & 1.58 & 14 & 3.02 & & 0.78 & 0.56 & 1.98 & &
5.77 & 0.85 &
25 \\
\textbf{4}& 11.2 & & 7.98 & 2.76 & 1.93 & ~9 & 3.30 & & 0.71 & 0.59 & 1.70 & &
5.95 & 0.89 &
25 \\
  \hline
\end{tabular}
\end{table*}

Having available the parameters of each model we consider as length of the bar
the length of the semi-major axis of the Ferrers bar, $a$. Also, from the
pattern speed of each model, $\Omega_b$, we compute the location of the
Lagrangian points $L_1$ and $L_2$. We consider as corotation radius their
distance from the center of the system. Then we calculated the ratio
$R=R_{CR}/a_{b}=R_{L_1}/a$. For the three models we analyzed, the values we
found are given in Table~\ref{tabtwo}.
\begin{table*}
\centering \footnotesize \caption{Parameters associated with the pattern 
speed 
of the studied models. Each row gives successively name of the snapshot, the 
time after the beginning of the simulation it is taken, the radius of the 
Lagrangian point $L_1$ and the corotation-to-bar's length ratio.}
\label{tabtwo}
\begin{tabular}{l l l l}
\hline
snapshot 2    & $t=4.25$  Gyr & $R_{L_1}=10.83$ & $R=2.0$ \\
snapshot 3    & $t=7.00$  Gyr & $R_{L_1}=16.46$ & $R=2.3$ \\
snapshot 4    & $t=11.2$  Gyr & $R_{L_1}=22.88$ & $R=2.87$ \\
  \hline
\end{tabular}
\end{table*}

We note that the $R_{L_4}$ values are very close o the $R_{L_1}$ ones, being 
10.75, 16.37 and 22.89 respectively. We realize that in all cases $R>1.4$, which 
places all models to the class of slow rotators.

\section[]{Autonomous Hamiltonian system and the GALI$_2$ index}
\label{sec:Autonomous Hamiltonian system and the GALI$_2$ index}
The Hamiltonian of the system is given by
\begin{equation}
\label{eq:h}
 H =
\frac{1}{2}(p_{x}^2+p_{y}^2+p_{z}^2)+\Phi(x,y,z)-\Omega_{b}(xp_{y}-yp_{x})=E_{J}
\end{equation}
where $x,y,z$ are Cartesian coordinates, $p_{x}, p_{y}, p_{z}$ are the 
conjugate momenta in the inertial reference frame, and $\Omega_{b}$ is the pattern 
speed of the bar. $E_{J}$ is the energy of Jacobi and $\Phi=\Phi_B + \Phi_D + 
\Phi_H$. where $\Phi_B$ is the potential of the bar, $\Phi_D$ is the potential of the disk,
and $\Phi_H$ is the potential of the halo.

The equations of motion and the variational equations we use in order to 
follow the evolution of the two deviation vectors from the studied orbit can be 
found in the MM paper. They are needed for computing the GALI$_2$ index.

The GALI$_2$ index is given by the absolute value of the wedge product of two
normalized to unity deviation vectors $\mathbf{\hat{w}}_{1}(t)$ and
$\mathbf{\hat{w}}_{2}(t)$:
\begin{equation}
 \rm{GALI}_{2}(t)=|\mathbf{\hat{w}}_{1}(t)\wedge \mathbf{\hat{w}}_{2}(t)|.
\end{equation}
(see MM for details). 

Thus, in order to evaluate GALI$_2$  we integrate the equations of motion and
the variational equations for two deviation vectors simultaneously.
The GALI$_2$ index behaves as follows \citep[see][and references 
therein]{2016Springer}:
\begin{itemize}
\item For chaotic orbits it falls exponentially to zero as:
\begin{equation}
 \mbox{GALI}_{2}(t)\propto \exp{(-(\lambda_{1}-\lambda_{2})t)}
\end{equation}
where $\lambda_{1}$ and $\lambda_{2}$ are the two largest Lyapunov exponents
\citep[for the computation of the Lyapunov exponents 
see:][]{1980Mecc...15....9B, 2010LNP...790...63S}.
\item For regular orbits it oscillates around a positive value across the
integration:
\begin{equation}
 \mbox{GALI}_{2}(t)\propto constant.
\end{equation}
\item 
In the case of sticky orbits we observe a transition from practically
constant GALI$_2$ values, which correspond to the seemingly quasi-periodic
epoch of the orbit, to an exponential decay to zero, which indicates the
orbit's transition to chaoticity.
\end{itemize}


\section[]{The degree of chaoticity of the orbits}
\label{degchaos}
\subsection[]{Planar orbits}
\label{2dchaos}
In a rotating Ferrers bar the elliptical periodic
orbits
of the main families are characterized by
a single non-zero initial condition along the minor axis of the bar, namely
their position along the y-axis in our models. The curve of zero velocity (ZVC)
in a $(E_J,y_0)$ diagram separates the region where orbital motion is allowed
from the region where it is not. Since the main family consists of direct
periodic orbits, only the $y_0>0$ part of such a diagram is of interest for us.
An $(E_J,y_0)$ diagram is the projection of a complete
$(E_J,y_0,p_{y_0})$
figure with all possible initial conditions. However, it is sufficient for
describing the properties of the orbits we present below.
The line that gives the $y_0$ initial condition of the main family of periodic
orbits is the characteristic curve of the model. Since we want to study
chaoticity in a large range of energies, we have created such
$(E_J,y_0)$ diagrams for the potentials of the three snapshots we study. In
order to calculate the degree of chaoticity of the \textit{planar} orbits around
the main family of periodic orbit as we move from the center of the system
towards corotation, we use the GALI$_2$ index. We have used the GALI$_2$ index
to color-code each point in the allowed region in the $(E_J,y_0)$ areas. The
shade of the color\footnote{In the electronic version of the paper we use
shades of blue to
colour-code the orbits. However, in the printed version the corresponding
figures are given in shades of gray.} indicates the GALI$_2$ index that a given
orbit, i.e. a point in the $(E_J,y_0)$ diagram, has at the end of the
integration. In other words, the
color of an $(E_J,y_0)$ point indicates if the orbit with $y_0$ initial 
condition at $E_J$ will lead to
regular (large $\log_{10}(\mbox{GALI}_2)$ values) or chaotic (very small
$\log_{10}(\mbox{GALI}_2)$ values) motion. At the borders between these
regions we find points with intermediate $\log_{10}(\mbox{GALI}_2)$
values, which correspond to sticky chaotic orbits.

\subsubsection{Snapshot 2, t=4.2 Gyr}
\label{sec:s2}
\begin{figure*} 
\begin{center}
\includegraphics[scale=0.60]{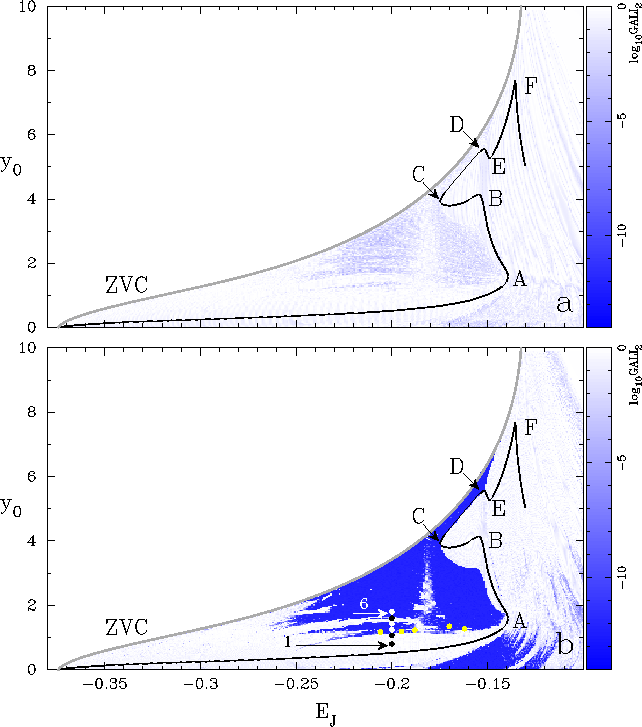}\\
\end{center}
    \caption{The chaoticity of the planar orbits on the equatorial plane of 
model ``2'' is given color coded in $(E_J,y_0)$ diagrams. The colour of each
orbit (each point in the figures) corresponds to the value of the
$\log_{10}$(GALI$_2$) quantity calculated for it and is taken from the
color-bar on the right hand side of the figures. In (a) we calculate
$\log_{10}$(GALI$_2$) for $t_1=$~1~Gyr, while in (b) for $t_2=$~10~Gyr. In both
figures the zero velocity curve is indicated with ``ZVC''. The continuous black
line in the region where motion is allowed is the characteristic of the main
family. Capital letters (A,B,...F) and arrows pointing to the points ``C'' and
``D'' are used for facilitating the description of the evolution of the curve in
the text. We observe that in general the orbits with the smaller GALI$_2$ index
in (a), which reach values $\log_{10}$(GALI$_2$)$\lessapprox -5$, become
strongly chaotic in (b). However, in (b) appear also additional features
indicating chaotic behavior, that are absent in (a). Such features are the dark
blue tails above the characteristic for $-0.27\lessapprox E_J \lessapprox
-0.17$. The six heavy dots at $E_J=-0.2$ indicate the initial conditions of the
orbits we use to demonstrate the relation between GALI$_2$ and their morphology
in Fig.~\ref{forbs}. Arrows point to the 1st and 6th of them. The five heavy,
light gray (yellow in the on-line version), dots at $E_J = -0.206, -0.195,
-0.18798, -0.17$ and $-0.162$ indicate the initial conditions of the boxy orbits
we present in Fig.~\ref{fiveo}.}
    \label{s2ej}
\end{figure*}

For each model we sample the GALI$_2$ index at two time windows. First after
time
$t_1$, corresponding to 1~Gyr and then after time $t_2$, corresponding to
10~Gyr. In this way we investigate both the relatively short-term as well as the
long-term behaviour of the orbits. The two color-coded $(E_J,y_0)$ diagrams
for snapshot ``2'' are given in Fig.~\ref{s2ej}. Fig.~\ref{s2ej}a gives the
index after $t_1=$~1~Gyr and Fig.~\ref{s2ej}b after $t_2=$~10~Gyr.
Darker shades indicate more chaotic orbits. The color for each orbit is
determined according to its $\log_{10}$(GALI$_2$) value and is taken from the
colour bars given to the right of the figures. 

In Fig.~\ref{s2ej} and all similar subsequent figures, the curve of zero
velocity is indicated with ``ZVC''. As determined by Eq.~\ref{eq:h}, motion is
allowed only to the right of the ZVC as drawn in Fig.~\ref{s2ej}. The local
$E_J$ maximum of the ZVC to the right of the figure gives the location of the
Lagrangian point $L_4$. The continuous heavy black curve in the region where
motion is allowed is the characteristic of the main family of periodic orbits.
We observe that it does not grow monotonically from the center towards
corotation, but after reaching point A at $E_J\approx -0.139$ it turns
backwards, towards lower energies. After reaching a local maximum in point B, it
changes again direction at $E_J\approx -0.175$ building a conspicuous open loop
in the energy range $-0.175\lessapprox E_J \lessapprox -0.152$. The loop becomes
evident by following the points B, C, D and E. The turn back of the
characteristic of the main family resembles the one encountered in the 2D model
in \citet{2015MNRAS.448.3081T} as well as the one of model ``A2'' in
\citet{2002MNRAS.333..861S}. The joining of x1-, x2- and x3-like morphologies in
a single, continuous characteristic, has been called by
\citet{1989A.ARv...1..261C} a ``type 4 gap''. 

Following the morphological evolution of the periodic orbits along this 
characteristic we realize that along its lower branch, for $E_J\lessapprox 
-0.139$, as well as between A and B, i.e. from $E_J\approx -0.139$ to 
$E_J\approx -0.155$ they are elliptical, extending along the bar. However, only 
the orbits with $E_J\approx -0.139$ match the size of the bar of the model. 
Between A and B we find ellipses larger than the bar, as it is indicated in 
figure 1 of MM. This means that such orbits are not populated in the model. 
Then, along the open loop the ellipticity of the orbits decreases. They become 
circular and then again elliptical, but extending this time along the minor axis 
of the bar, i.e. they are x2-like. For $E_J\gtrapprox -0.152$, the periodic 
orbits of the main family are almost circular. We do not include in 
Fig.~\ref{s2ej} the characteristics of $n:1$--resonance families with $n\geq 4$ 
beyond the gap at the radial 4:1 resonance. In this paper we are interested in 
orbits supporting boxiness and the bar supporting otbits close to corotation are 
practically planar \citep{2002MNRAS.333..861S} with circular projections on the 
equatorial plane.

In Fig.~\ref{s2ej}a color characterizes the chaoticity of the orbits after
integrating them for $t_1=1$~Gyr. Within this time it is expected that not only
regular but also weakly chaotic, sticky orbits will retain a regular character.
Such orbits will be able to support a given structure during this time period.
Keeping the same scale in the coloured bars on the right hand sides of the
figures we can compare the evolution of the chaotic character of the orbits
from Fig.~\ref{s2ej}a to Fig.~\ref{s2ej}b. The same shade indicates the same
degree of chaoticity in the two figures. 

Fig.~\ref{s2ej}b, gives the same information with the only difference that the 
time of integration is $t_2=10$~Gyr. There is an overall similarity between the 
two figures. The orbits with the larger GALI$_2$ values in Fig.\ref{s2ej}a 
(light blue areas) developed a clear chaotic character within $t_2$ (dark blue 
areas in Fig.~\ref{s2ej}b). We also observe that there is a white stripe 
surrounding the characteristic of the main family almost for all energies in 
both figures. This indicates that the periodic orbits of the main family are 
stable and thus small perturbations of their initial conditions lead to regular 
motion characterized by large $\log_{10}$(GALI$_2$). A notable exception is the 
region between the CD part of the characteristic and the ZVC in 
Fig.~\ref{s2ej}b. A very thin dark layer exists also just below this 
part of the characteristic. We remind that along the same characteristic curve 
of the main family of our models we encounter morphologies of periodic orbits 
that correspond to the stable families x1 and x2, but also to the unstable x3 
family. In other models these three families have disconnected characteristics 
\citep[see e.g.][]{1989A.ARv...1..261C, 1992MNRAS.259..328A, 
2014MNRAS.445.3525P}. We also observe that in Fig.~\ref{s2ej}b there are clearly 
developed dark blue tails with chaotic orbits, absent in Fig.~\ref{s2ej}a, in 
the region above the characteristic of the main family for $-0.27\lessapprox E_J 
\lessapprox -0.17$. Another conspicuous white zone extends almost perpendicular 
to the $E_j$ axis at about $E_J\approx -0.18$. It shrinks when we integrate the 
orbits for $t_2=10$~Gyr (Fig.~\ref{s2ej}b). 

We consider now orbits along a line of constant $E_J$ in Fig.~\ref{s2ej}b at
which regular and chaotic regions alternate, in order to investigate the
morphology-GALI$_2$ relation. Such an energy is for example $E_J=-0.2$. We
observe that along the $E_J=-0.2$ axis we encounter both regular and chaotic
regions, depicted as a succession of blue (chaotic) and white (regular) regions.
We present the behavior of six planar orbits at this energy with initial
conditions $y_0 = 0.80, 1.07, 1.24, 1.43, 1.60$ and $1.80$. In all cases
$p_{y_0}=0$ (we remind that the major axis of the bar is along the x-axis of our
system). We name these orbits ``1'', ``2'', ...,``6'' and denote their location
in Fig.~\ref{s2ej}b with six heavy dots. We use black or white heavy dots
depending on the background in order to make them as discernible as possible.
Arrows point to the location of the first (``1'') and sixth (``6'') of these
orbits. Moving along a line of constant energy we obtain some of the information
a Poincar\'{e} surface of section provides. GALI$_2$ reveals the succession
of regular and chaotic motion along the $p_{y_0}=0$ axis in the Poincar\'{e}
section at this energy. The width of the white space on both
sides of the characteristic of the main family at a given energy is associated
with the size of the stability island around the stable periodic orbit. The
crossing of white stripes by an axis of constant $E_J$ corresponds to other,
smaller, islands of stability that exist on the surface of section $p_{y_0}=0$.

\begin{figure*} 
\begin{center}
\includegraphics[scale=0.6]{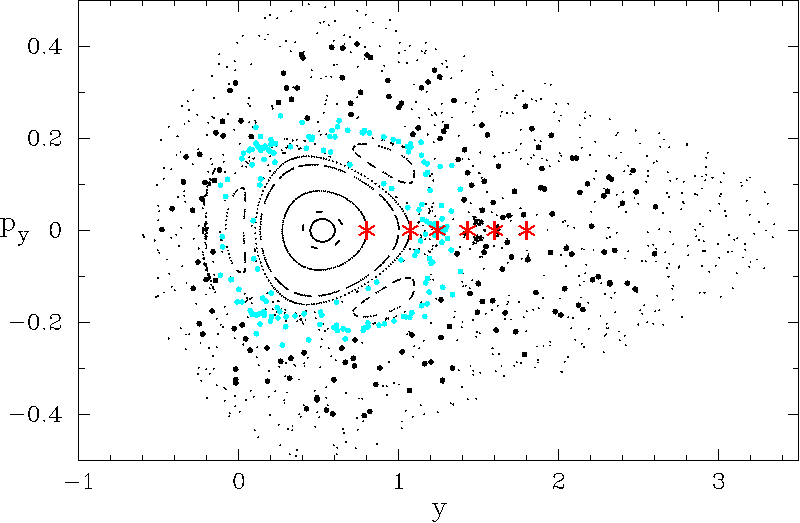}\\
\end{center}
    \caption{The Poincar\'{e} surface of section of model 2 for
    $E_J=-0.2$. The stable periodic orbit at $(y,p_y)\approx(0.5,0)$
    belongs to the main family of planar periodic orbits of the
    system. The 6 big asterisks indicate the initial conditions of the
    orbits labeled with ``1'' to ``6'' in Fig.~\ref{s2ej}b.}
    \label{sos2}
\end{figure*}

\begin{figure*} 
\begin{center}
\includegraphics[scale=0.70]{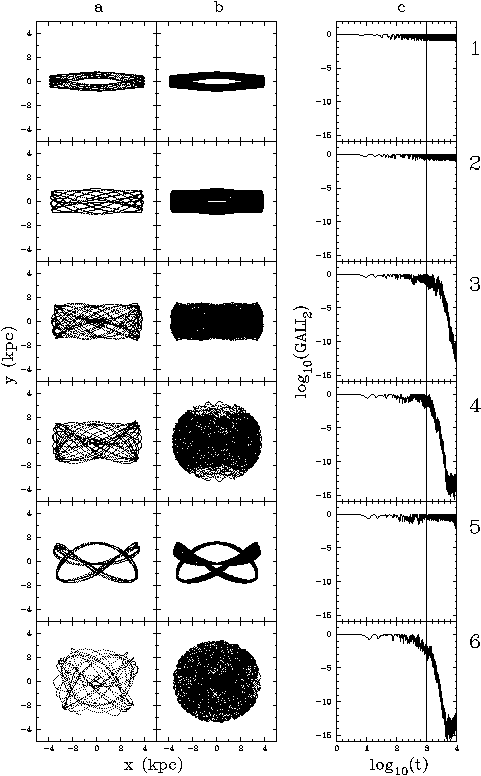}\\
\end{center}
    \caption{The six orbits with initial conditions indicated in
    Fig.~\ref{s2ej}b (``1'' to ``6'') and in Fig~\ref{sos2}
    (asterisks). Each row refers to the orbit mentioned at its right
    hand side. The columns (given above them) show: (a) The morphology
    of the orbit within $t_1=1$~Gyr, (b) The morphology
    of the orbit within $t_2=10$~Gyr, (c) The evolution of
    $\log_{10}$(GALI$_2$) within $t_2=10$~Gyr. The vertical line
indicates the location of the 1st Gyr.}
    \label{forbs}
\end{figure*}

In Fig.~\ref{sos2} we present the Poincar\'{e} surface of section for
$E_J=-0.2$. 
The six big asterisks along the $p_{y_0}=0$ axis with $0.8\leq y \leq
1.8$ are, from left to right, the initial conditions of the orbits
``1'' to ``6'' indicated in Fig.~\ref{s2ej}b. The evolution of the
morphologies and of the quantity $\log_{10}$(GALI$_2$) for these orbits
within $t_1=1$~Gyr and $t_2=10$~Gyr is given in Fig.~\ref{forbs}.

Orbit ``1'', with $y_0=0.8$ (the lowest initial condition indicated with ``1'' 
in Fig.~\ref{s2ej}b) corresponds to the left asterisk in Fig~\ref{sos2}. From 
its location in the surface of section we can see that it belongs to an 
invariant curve on the stability island around the stable representative of the 
main family of periodic orbits and close to it. At this energy the stable 
periodic orbit is a typical x1 ellipse. The quasi-periodic orbit we study has a 
morphology that can be vaguely described as a ``thick'' ellipse (panels a and b 
in row ``1'' in Fig.~\ref{forbs}). Since it is a regular orbit 
$\log_{10}$(GALI$_2$) fluctuates close to 0 (panel c in row ``1'') as expected.

Orbit ``2'' has $y_0=1.07$ and belongs also to an invariant curve. The
invariant curve around orbit ``2'' is very close to the last
KAM (Kolmogorov-Arnold-Moser) curve \citep[see e.g.][]{2002ocda.book.....C}, of
the main stability island of Fig~\ref{sos2}. As we observe in panel (c) of the
second row of Fig.~\ref{forbs}, also in this case
$\log_{10}$(GALI$_2$) fluctuates close to 0. The morphology of the orbit
is boxy (panels a and b for ``2''). However, we observe that even
after integration time that corresponds to 10~Gyr there is a central
region that is not visited by the orbit.

The next orbit, ``3'' (with $y_0=1.24$), gives the gray (light blue in the 
on-line version), heavy consequents in Fig~\ref{sos2}. For most of the 
integration time these consequents are trapped around three stability islands 
located just beyond the invariant curve of orbit ``2''. However, close to the 
end of the integration time, orbit ``3'' starts diffusing in the larger chaotic 
sea surrounding the region with the stability islands. Thus, it is a typical 
sticky orbit. The quantity $\log_{10}$(GALI$_2$) is very close to 0 during the 
first Gyr, reaching $-1$ close to the end of this time period (cf. location of 
vertical line in panel c in the third row in Fig.~\ref{forbs}). Beyond that time 
and up to 10~Gyr it clearly decreases revealing the chaotic character of the 
orbit (panel c in the third row). The morphology of the orbit is boxy both for 1 
as well as for 10~Gyr (panels a and b in row ``3''). Diffusion in configuration 
space is observed only during the time the consequents start visiting all the 
available space in the surface of section. However, in Fig~\ref{sos2}  we see 
that the light blue dots remain confined in a specific region. In panel (a) we 
observe that we have the formation of an X feature inside the box. The feature 
exists also in panel (b) of row ``3''.  We find that an orbit sticky to the 
stability island of an x1 periodic orbit has a boxy morphology with an X 
embedded in it. Thus, we have in this case the formation of a boxy orbit 
supporting an X feature on the galactic plane by means of the dynamical 
mechanism described by \citet{2015MNRAS.448.3081T}.

The orbit with $y_0=1.43$ (4th asterisk from left in Fig~\ref{sos2}) starts in
the chaotic sea. Its consequents are drawn with heavy black dots in the surface
of section. We observe that they are distributed in a larger region than the
consequents of orbit ``3'', while they almost do not visit the region occupied
by the light blue/gray consequents. For 1~Gyr it also supports a boxy bar with
an X feature as orbit ``3'' (panel a in the 4th row of Fig.~\ref{forbs}) having
also a rather regular behavior with $\log_{10}$(GALI$_2$) close to 0 (panel c,
before the vertical line for ``4''). However, for larger time,
$\log_{10}$(GALI$_2$) decreases abruptly reaching smaller values than the case
of orbit ``3'' (cf. panels c in 3rd and 4th row) and, contrarily to orbit ``3'',
 has a chaotic morphology (panel b in ``4''). We also observe that for orbit
``3'', after 10~Gyr $\log_{10}$(GALI$_2 \approx -12$), while this happens
already at $t\approx$ 5000 for orbit ``4''. 

Moving along the the $p_y=0$ axis towards larger $y$ in
Fig~\ref{sos2}, we enter a zone occupied by barely discernible  
stability
islands. Without going into details for
the periodic orbits we find
there, we just mention that in the region there is a periodic orbit of
multiplicity 6. This region corresponds
to the white stripe below the arrow labeled with ``6'' in
Fig.~\ref{s2ej}b. Orbit ``5'' ($y_0=1.6$) is \textit{almost} on the invariant
curves of the 6-ple orbit. The $\log_{10}$(GALI$_2$) index points to a
regular orbit (panel c in row ``5'' of Fig.~\ref{forbs}),
which is in agreement with the morphologies after 1 and 10~Gyr as we can 
observe in panels
a and b in ``5'' respectively. Actually this is also a sticky orbit whose 
chaotic nature will be revealed for $t > 10$~Gyr, as towards the end of the 
integration we can observe a gradual decrease of the $\log_{10}$(GALI$_2$) 
quantity.

Finally, starting with $y_0=1.8$ (orbit ``6'') we find a chaotic orbit
(scattered small dots in Fig~\ref{sos2}). The morphologies in panels (a)
and (b) in the 6th row of Fig.~\ref{forbs} and the corresponding
$\log_{10}$(GALI$_2$) index (panel c) are in agreement with the
chaotic nature of the orbit.

As we observe in column a of Fig.~\ref{forbs}, within $t_1=1$~Gyr the orbits
``1'' to ``5'' evidently support some structure. Only orbit ``6'' has a well
developed chaotic character. For larger time, $t_2=10$~Gyr, besides orbit ``6'',
also orbit ``4'' has a chaotic morphology (panels b in ``4'' and ``6'' in
Fig.~\ref{forbs}). The boxy orbital structures that we are looking for are not
encountered in all structure supporting orbits. The regular orbits ``1'' and
essentially ``5'' belong to invariant curves close to the initial conditions of
the periodic orbits and their morphology reflects to a large extent their
morphology. Clear boxiness appears in orbits ``2'' and ``3''. They are located
in the outermost parts of the stability island of the main periodic orbit (orbit
``2'') and in the sticky region around it (orbit ``3'') respectively, as we can
observe in Fig.~\ref{sos2}. Their regular and sticky behavior is reflected also
to their $\log_{10}$(GALI$_2$) index within $t_2=10$~Gyr (panels c of ``2'' and
``3'' in Fig.~\ref{forbs}). The chaotic orbit ``4'' has a morphology similar to
``3'' only during the first Gyr of integration. This result is in agreement with
the result of \citet{2015MNRAS.448.3081T}, namely that boxiness in face-on views
of bars at a given energy is introduced by orbits at the critical area close to
the last KAM curve around the stable x1 orbit. They can be either on the regular
or sticky side. In the latter case we have also the appearance of an embedded X
feature. 

\begin{figure*} 
\begin{center}
\includegraphics[scale=0.75]{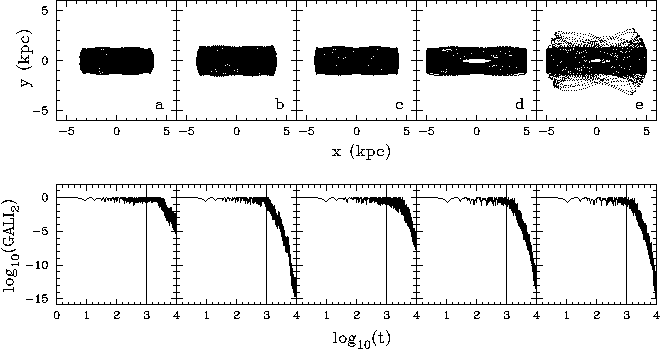}\\
\end{center}
    \caption{Five orbits with boxy character on the equatorial plane of the
model of snapshot 2. Their location on the $(E_J,y_0)$ diagram are denoted with
heavy light gray (yellow in the on-line version) dots in Fig.~\ref{s2ej}b. They
are: (a) $(-0.206,1.17)$, (b) $(-0.195,1.19)$, (c) $(-0.18798,1.23)$, (d)
$(-0.17,1.35)$ and (e) $(-0.162,1.27)$. All of them are sticky to the stability
islands of the stable representative of the main family of periodic orbits of
the system. Below its panel of the first row is given the corresponding GALI$_2$
index up to 10~Gyr. The vertical line indicates the location of the 1st Gyr.}
    \label{fiveo}
\end{figure*}

In order to demonstrate the relation between boxiness of the orbits and their
location close to the borderline between order and chaos, we considered five
more orbits in this zone at various energies. These are the orbits presented in
Fig.~\ref{fiveo}. 
All of them are sticky, located inside the dark area, but close to the
borderline between white and dark (blue in the on-line version) regions in
Fig.~\ref{s2ej}b. They are indicated with heavy light gray (yellow in the
on-line version) dots at the energies: $E_J = -0.206, -0.195, -0.18798, -0.17$
and $-0.162$. Their initial $y_0$ values are respectively 1.17, 1.19, 1.23, 1.35
and 1.27 (always with $p_{y_{0}}$=0). In Fig.~\ref{fiveo} we give them
successively from left to right with increasing energy. Below each panel with
the orbit in the $(x,y)$ plane we give its $\log_{10}$(GALI$_2$) index. As
GALI$_2$ shows, all five orbits  manifest their chaotic nature at times larger
than 1~Gyr (indicated in all lower panels with a vertical line). Orbits in
Fig.~\ref{fiveo}a to d, remain confined in the configuration space until the end
of the integration time, i.e. for 10~Gyr. The orbit in Figs.~\ref{fiveo}e, at
the largest energy, is more chaotic. It reaches a smaller
$\log_{10}$(GALI$_2$) value at time 10~Gyr, while, close to the end of
its integration time, it starts exploring larger regions in the
configuration space. However it also has a boxy morphology within time
corresponding to about 1~Gyr. 

The above results point out that in order to find orbits on the equatorial plane
that support boxy features in the bar of the model we have to consider initial
conditions close to the borderline between order and chaos in Fig.~\ref{s2ej}.
This happens not just close to a specific resonance. We find such orbits
at all energies where exist x1 periodic orbits matching the size of the bar.
The regions, where one should look for candidate orbits supporting boxy features
in the models, are of those which still appear white after 1~Gyr of integration
in Fig.~\ref{s2ej}a and are found being marginally inside the chaotic region in
Fig.~\ref{s2ej}b, i.e. have developed a chaotic character in time $1<t<10$~Gyr.

\subsubsection{Snapshot 3, t=7.0 Gyr}
\begin{figure*} 
\begin{center}
\includegraphics[scale=0.70]{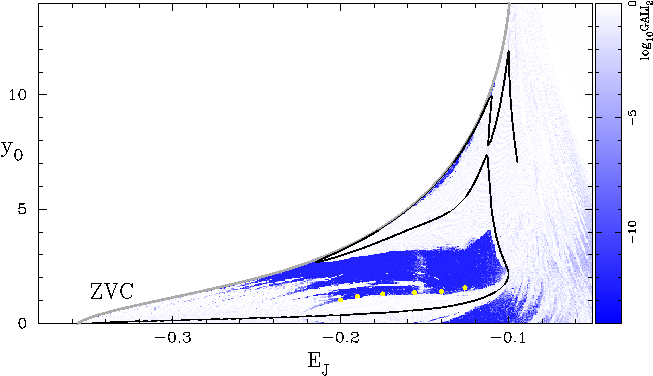}\\
\end{center}
    \caption{The chaoticity of the planar orbits on the equatorial plane of
model ``3''. The drawn lines and the given colors are as in Fig.~\ref{s2ej},
which is the corresponding figure for model ``2''.  Here we calculate
$\log_{10}$(GALI$_2$) for $t_2=$~10~Gyr. The six heavy, light gray (yellow in
the
online version), dots indicate the initial conditions of the boxy orbits we
present in Fig.~\ref{sixorb3}.}
    \label{s3ej}
\end{figure*}

\begin{figure*} 
\begin{center}
\includegraphics[scale=0.80]{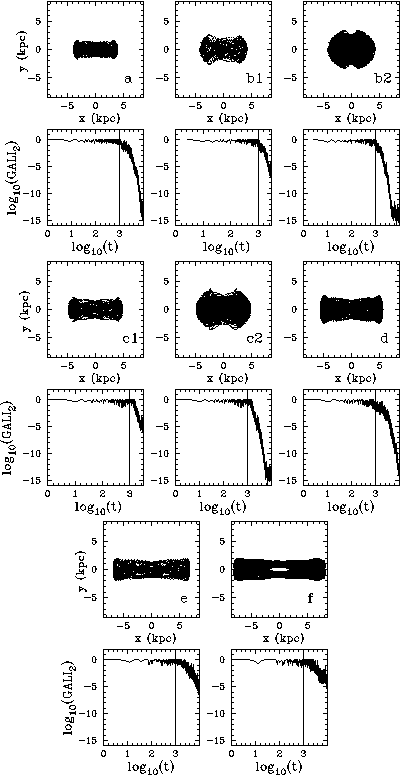}\\
\end{center}
    \caption{Six orbits with boxy character on the equatorial plane of the
model of snapshot 3. Their location on the $(E_J,y_0)$ diagram are denoted with
heavy light grey (yellow in the on-line version) dots in Fig.~\ref{s3ej}. They
are: (a) $(-0.2,1.03)$, (b) $(-0.19,1.19)$, (c) $(-0.175,1.3)$, (d)
$(-0.156,1.35)$, (e) $(-0.14,1.4)$ and (f) $(-0.126,1.57)$. For orbits
in (b) and
(c) we give their morphology in two different time windows (b1, b2 and c1, c2
respectively). All of them are sticky to the stability
islands of the stable representative of the main family of periodic orbits of
the system. Below each labeled panel is given the corresponding GALI$_2$
index up to 10~Gyr. The vertical line indicates the location of the 1st Gyr.}
    \label{sixorb3}
\end{figure*}

Then we repeat the same analysis for the model of snapshot 3. Figure \ref{s3ej},
like
Fig.~\ref{s2ej}, gives
colour-coded the $\log_{10}$(GALI$_2$) quantity in a $(E_J,y_0)$ diagram,
however this time only for
$t_2=$~10~Gyr. Boxy orbits are again found at the border line between order
and chaos. We present six of them in Fig.~\ref{sixorb3}. Their locations in 
Fig.~\ref{s3ej} are at $(E_J,y_0)= (-0.2,1.03), (-0.19,1.19), (-0.175,1.3), 
(-0.156,1.35)$, $(-0.14,1.4)$ and $(-0.126,1.57)$. The time evolution of the 
quantity $\log_{10}$(GALI$_2$) below each panel with the morphology of the orbit 
in Fig.~\ref{sixorb3}, implies that the presented orbits are sticky. 
them remain confined in the 
For $t<t_1 =1$~Gyr these orbits can hardly be distinguished from regular orbits. 
We pay special attention to the orbits of panels (b) and (c) that we present in 
two different time windows and so they are labeled b1, b2 and c1, c2 
respectively. In the first time window, which is larger than 1~Gyr, we plot the 
orbits for the time they are retaining its boxiness, while for the second one we 
plot the orbits as they appear after being integrated for $t_2=$10~Gyr. We 
observe that the orbit in b1 is boxy and harbors evidently an X structure. 
However, within 10~Gyr it shows a strongly chaotic character. Its morphology in 
the configuration space is  chaotic (panel b2) and GALI$_2$, below it, has a 
steep gradient downwards reaching values close to $10^{-14}$. We encounter a 
similar evolution in the orbit described in panels c1 and c2. In this case the 
orbit remains confined in the configuration space for more than 3~Gyr and then 
expands into a larger region of the configuration space, without however 
visiting for time 10~Gyr all the allowed space. 

We also note that the boxy orbits presented in Fig.~\ref{sixorb3} have along
the x=0 axis a clear $|y_{min}|$ value that gives them a bow-like
shape,
something that has its counterpart in the shape of the $N-$body bar in the MM
models (cf. figure 1, third panel from left, in MM). This will be
further discussed also
in Sect.~\ref{sec:conclusions}.
 
\subsubsection{Snapshot 4, t=11.2 Gyr}
\begin{figure*} 
\begin{center}
\includegraphics[scale=0.70]{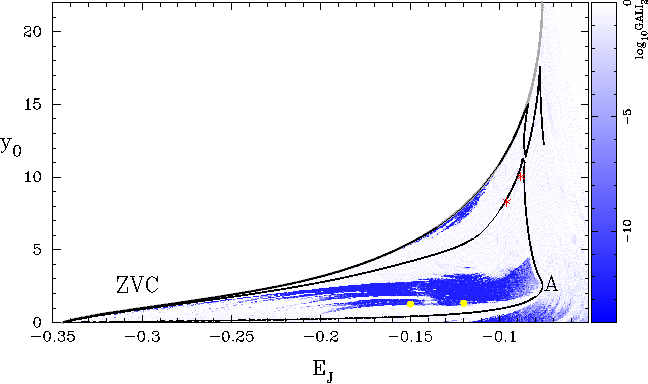}\\
\end{center}
    \caption{The chaoticity of the planar orbits on the equatorial plane of
model ``4''. The drawn lines and the given colors are as in Fig.~\ref{s2ej}.
Here we calculate $\log_{10}$(GALI$_2$) for $t_2=$~10~Gyr. The two heavy, light
gray (yellow in the on-line version), dots indicate the positions of orbits (a)
and (b) in Fig.~\ref{boxring4}, while the two asterisks, those of orbits (c) and
(d) in the same figure.}
    \label{s4ej}
\end{figure*}

\begin{figure*} 
\begin{center}
\includegraphics[scale=0.6]{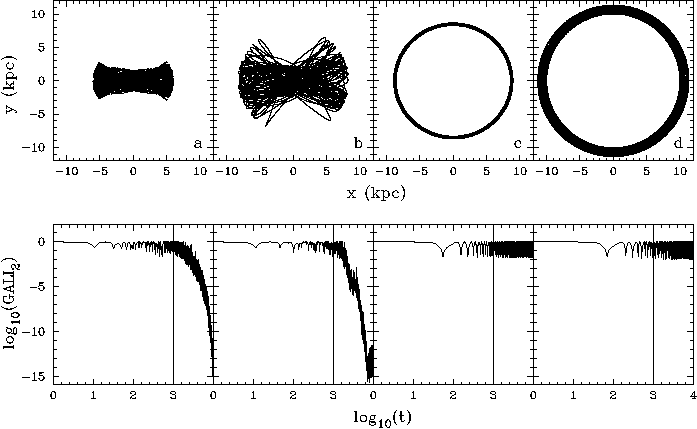}\\
\end{center}
    \caption{Four orbits  on the equatorial plane of the model of snapshot 4.
Their location on the $(E_J,y_0)$ diagram are denoted with heavy light grey
(yellow in the on-line version) dots  and asterisks in Fig.~\ref{s4ej}. They 
are:
(a) and (b) the orbits with the heavy dots located at $(E_J,y_0)=(-0.15,1.27)$
and $(-0.12,1.33)$ respectively and (c) and (d) the orbits with the two
asterisks at $(-0.096,8.292)$ and $(-0.088,10.032)$ in Fig.~\ref{s4ej}. Below
each labeled panel is given the corresponding GALI$_2$ index up to 10~Gyr. The
vertical line indicates the time corresponding to the 1st Gyr.}
    \label{boxring4}
\end{figure*}

Finally we repeat the same analysis for the last snapshot of the MM paper. The
colour-coded $(E_J,y_0)$ diagram for this case is Fig.~\ref{s4ej}.
This is a very slowly rotating model with $R=2.87$ and corotation at 22.89~kpc.
As we can see, the loop of the characteristic of the central family is huge. The
characteristic increases monotonically until point A and then turns backwards.
The branch that goes back to the left reaches the minimum $E_J$ of the
ZVC. Then it turns back again towards corotation, being essentially on the ZVC.
The loop almost closes as the two parts of the characteristic come very close at
about $E_J=-0.09$. Bar supporting orbits on the equatorial plane can be found
only in the lowest branch of the characteristic, while there is a large amount
of almost circular and stable orbits (practically white regions for $y_0> 5$ in
Fig.~\ref{s4ej}) that populate the extended disk region between the end of the
bar and corotation (cf. figure 1, right panel, in MM). Four typical orbits for
this model are given in Fig.~\ref{boxring4}. Their locations in the $(E_J,y_0)$
diagram are denoted with heavy light gray (yellow in the on-line version) dots
and with asterisks in Fig.~\ref{s4ej}. They are at: (a) $(-0.15,1.27)$, (b)
$(-0.12,1.33)$ (heavy dots), (c) $(-0.096,8.292)$, (d) $(-0.088,10.032)$
(asterisks).
The accumulation of a large number of almost circular orbits for $y_0$ initial
conditions beyond those of the bar supporting orbits and the shape of the
characteristic with the almost closed loop, favor the formation of rings
surrounding the bar by means of a dynamical mechanism similar to the one that
led to the formation of the ring in the model of \citet{2015MNRAS.448.3081T}.

\subsection{Vertical perturbations}
Until now we have seen that a set of planar orbits with boxy morphology can be 
found close to the border line between order and chaos above the characteristic 
of the central family of periodic orbits, as this is determined by the GALI$_2$ 
index in the $(E_J,y_0)$ diagrams. Now we will examine how the morphology of 
these orbits changes if we perturb them vertically by adding a $p_{z_0}\neq 0$ 
to their initial conditions. This means that the orbits we present in this 
section have initial conditions $y_0$ and $p_{z_0}\neq 0$, while $z_0$ and 
$p_{y_0}=0$. Hereafter, when we give the initial conditions of an orbit, we will 
mean the non-zero ones, if not otherwise indicated.

For 3D orbits their regular or chaotic character cannot be easily depicted on a
single diagram, since we deal in general with four initial conditions.
Considering an orbit, the monotonic variation of a single initial condition may
lead to a non-monotonic succession of regular and sticky chaotic orbits. It is
not easy to know in a 4D space whether the deviation from the initial condition
of a torus will bring an orbit in a chaotic sea or closer to an invariant torus
around another stable periodic orbit. However, we realized that for all planar
boxy orbits we started increasing their $p_{z_0}$ coordinate, we could find a
$\Delta p_z$ range for which the 3D orbits retained their boxy
character. The variation of the GALI$_2$ index with time was similar to
that of the boxy 2D orbits. This led to the conclusion that the building blocks
not only for 2D, but also for 3D boxy structures in real galaxies can be either
regular orbits on the most remote tori around stable periodic orbits, or orbits
sticky to them. The latter orbits are strictly speaking chaotic, but their
sticky character keeps them confined in particular regions of the phase space
for sufficiently long times. This way they can support a given morphology.

\begin{figure*} 
\begin{center}
\includegraphics[scale=0.75]{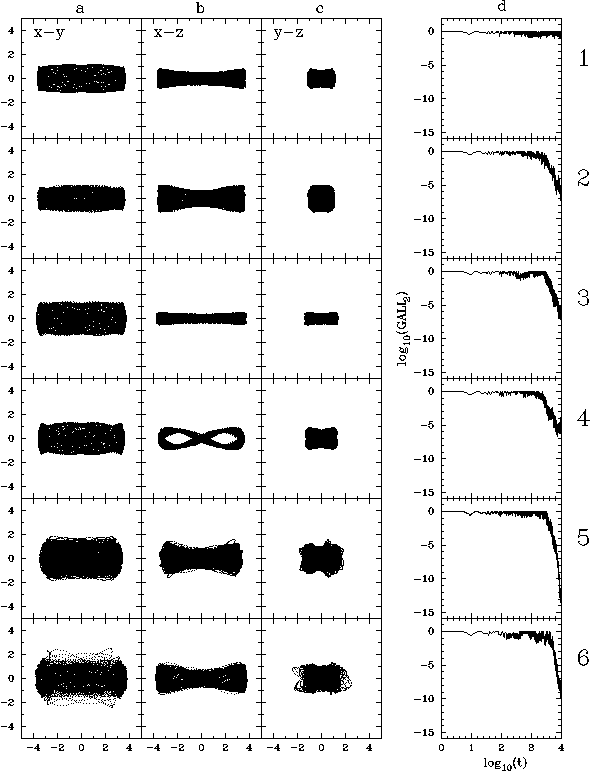}\\
\end{center}
    \caption{3D orbits associated with 3D boxy structures in the model of 
snapshot 2. Orbit ``4'' has a side-on profile similar to x1v2 orbits, while all 
the rest have three boxy projections. Orbits ``1'', ``3'', ``4'' and ``5'' 
harbour an X feature in their face-on projections. The GALI$_2$ evolution 
indicates the sticky character of the orbits ``2'' to ``6'', while ``1'' is 
regular. The units on the axes of the three first columns are in kpc.}
    \label{m2threed}
\end{figure*}

Below we give some typical examples of 3D boxy orbits, or, in other words, 
orbits with three boxy projections in the configuration space. In 
Fig.~\ref{m2threed} we present six orbits from the model of snapshot 2. The 
panels of each row refer to the same orbit. A number that refers to each orbit 
and helps us in the description, is given at the right hand side of each row. In 
column (a) we give the face-on views, in (b) the side-on view, in (c) the end-on 
one and finally in (d) the evolution of GALI$_2$ in a log-log plot as in the 
previous figures.

In general, by starting from a planar orbit and adding $p_z$ we have the 
following possibilities: 1. We will either reach a torus around x1, or 2. a 
torus around a stable 3D family bifurcated from x1, or 3. a chaotic zone between 
the two sets of tori, or finally 4. we will enter a chaotic zone \citep[see][and 
Patsis \& Harsoula 2017 -- in preparation]{2014MNRAS.445.3525P}. The result 
depends both on the energy of the orbit and the $p_z$ initial condition. The 
energy will determine the available resonant families of periodic orbit existing 
(their number increases with $E_J$), while $p_z$ will decide about the location 
of the orbit in the phase space. In the present paper we are interested just in 
pointing out that there are vertical perturbations that support the 3D boxy 
character. The planar orbits we start from have to be sought along the lines we 
find the 2D boxy orbits in the $(E_J,y_0)$ diagrams. In Fig.~\ref{m2threed}, the 
orbits in the four first rows are at the same energy we have the $(y,p_y)$ 
surface of section in Fig.~\ref{sos2}, i.e. for $E_J=-0.2$. The orbits ``1'' and 
``2'' have $y_0=1.07$, which would be an initial condition on an invariant curve 
around x1 in the $(y,p_y)$ Poincar\'{e} surface of section (Fig.~\ref{sos2}) if 
we had $p_z=0$. However, orbit ``1'' is perturbed by $p_z=0.15$ and orbit ``2'' 
by $p_z=0.2$. In both cases the orbits form boxes in all three projections. The 
side-on and end-on morphologies clearly support peanut-shaped structures. The 
GALI$_2$ evolution of orbit ``1'' (panel d) indicates that it is a regular 
orbit, while that of orbit ``2'' points to a sticky one. The next orbit, 
perturbed by $p_z=0.085$, has $y_0=1.24$, i.e. in the $(y,p_y)$ Poincar\'{e} 
surface of section in Fig.~\ref{sos2} corresponds to the sticky orbit plotted 
with the heavy gray/light blue consequents. Again in this case the 3D boxy 
character is retained, however this time the vertical perturbation is smaller. 
For the same energy we give an example of an orbit with $y_0=1.19$ and 
$p_z=0.165$, which is orbit ``4'' in Fig.~\ref{m2threed}. If $p_z=0$ the orbit 
would be a sticky to x1 orbit. Now it is sticky again, but its morphology 
clearly resembles the morphology of the x1v2 family, which is bifurcating, 
usually as unstable, at the vertical 2/1 resonance \citep{2002MNRAS.333..847S}. 
This can be seen in panel (b) of row ``4'' in Fig.~\ref{m2threed}. Strictly 
speaking this side-on profile is not boxy. Nevertheless it has a shape similar 
to the one of the two main vertical bifurcations of x1. Considering several 
orbits like this in different energies will lead to a boxy profile. The orbits 
``5'' and ``6'' are in nearby energies and have similar morphologies and 
evolution of GALI$_2$ as the previous ones. Orbit ``5'' is at $E_J=-0.206$ with 
$y_0=1.17$ and $p_z=0.162$, while orbit ``6'' at $E_J=-0.195$ with $y_0=1.19$ 
and $p_z=0.15$. 

\begin{figure*} 
\begin{center}
\includegraphics[scale=0.75]{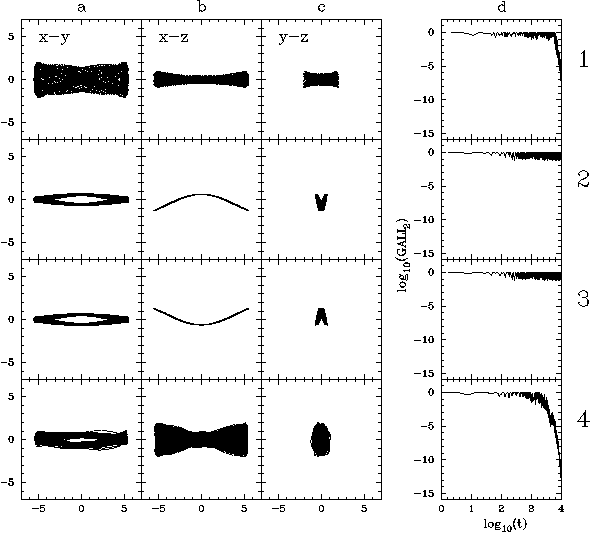}\\
\end{center}
    \caption{Orbits in the model of snapshot 3. ``1'' Sticky orbit with a boxy 
3D structure and an X feature embedded in the face-on projection, ``2'' and 
``3'' frown and smile regular 3D orbits on x1v1 tori, ``4'' a sticky orbit that 
changes its face-on elliptical morphology becoming boxy as soon as it abandons 
its regular behavior. The units on the axes of the three first columns are in 
kpc.
}
    \label{m3threed}
\end{figure*}

An interesting result is that the sticky 3D boxy orbits in many cases harbor an
X feature in their \textit{face-on} projections (column a). This is conspicuous
in orbits ``3'', ``4'' and ``5'', as well as in the regular orbit ``1''. This is
in agreement with the result of \citet{2014MNRAS.445.3546P} who suggest that
sticky boxy orbits at the immediate neighborhood of the vertical 2:1
resonance have embedded X features in their face-on projections. The property
of stickiness was also the reason for the appearance of an X inside the bars of
the 2D barred-spiral models in \citet{2015MNRAS.448.3081T}. The same analysis
led to similar results in the cases of the two other models considered here as
well. In Fig.~\ref{m3threed} we present some typical orbits for the model of
snapshot 3. Again here, double boxiness with an X feature embedded in the boxy
face-on projection is found by perturbing boxy planar orbits in the z direction
by $p_z$. This is given in orbit ``1'', which is for $E_J=-0.156$, $y_0=1.35$
and $p_z=0.09$. In Fig.~\ref{sixorb3}d we have given the corresponding orbit
with $p_z=0$.  As the orbits ``2'' and ``3'' show, the stable 3D families
\citep[x1v1 and x1v1$^{\prime}$ in the notation of][]{2002MNRAS.333..847S}
bifurcated from x1 
at the vertical 2/1 resonance, do exist in the model. In order to track them we
perturbed in the vertical direction the z coordinate, while we put the initial
condition $p_z=0$. The initial conditions of the two orbits are $E_J=-0.156$,
$y_0=0.7$ and $z=0.63$ (orbit ``2'') and $E_J=-0.156$, $y_0=0.7$ and $z=-0.63$
(orbit ``3'') respectively. Their morphology indicates that they belong to
invariant tori \textit{in the immediate neighborhood} of x1v1
\citep{2014MNRAS.445.3546P}. This is consistent with the evolution of their
GALI$_2$ index in panels (d). By means of such orbits we can construct
a sharp X-shaped side-on profile, having however an elliptical face-on
morphology. A nice example is given with orbit ``4'' with non-zero initial
conditions $E_J=-0.156$, $y_0=0.7$ and $p_z=0.303$. This is a sticky orbit (see
panel d). As long as it has a regular character, the orbit supports an
elliptical face-on morphology. However, when it starts diffusing in the
configuration space it tends to obtain a boxy structure (panel a). 

\begin{figure*} 
\begin{center}
\includegraphics[scale=0.75]{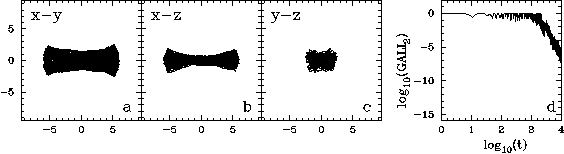}\\
\end{center}
    \caption{A 3D boxy orbit in the model of snapshot 4. It is sticky (panel d)
and reproduces the X feature in its face-on projection (panel a). The units on
the axes of the three first panels are in kpc.}
    \label{m4threed}
\end{figure*}

Finally in Fig.~\ref{m4threed} we give an example of an orbit from the model of 
snapshot 4, that reproduces the main morphology we want to underline that
exists in our models. Namely, it is a sticky orbit with all its projections
boxy, while in its face-on projection it is discernible an X feature. The
initial conditions of the orbit are: $E_J=-0.15$, $y_0=1.27$ and $p_z=0.07$.

\begin{figure*} 
\begin{center}
\includegraphics[scale=0.75]{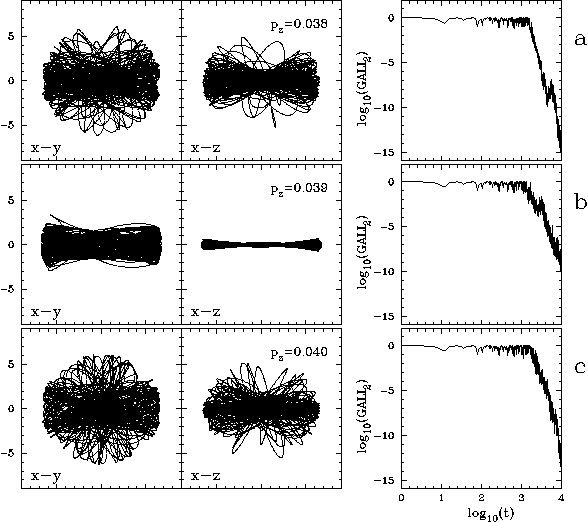}\\
\end{center}
    \caption{The planar orbit presented in Fig.~\ref{sixorb3}e perturbed by
$p_z=0.038$ (a), $0.039$ (b) and $0.040$ (c). In (a) and (c) the orbits diffuse
in configuration space, while in (b) the orbit has for more than 3~Gyr a double
boxy character.}
    \label{nonmon}
\end{figure*}

The process for finding 3D double boxy orbits by perturbing 2D boxy ones
can be applied at all energies, for which we could find x1 orbits supporting the
size of the $N$-body bar in the models. However, as $E_J$ increases and we
approach corotation, the structure of phase-space becomes more complicated, due
to the presence of more families of periodic orbits introduced in the system at
successive resonances \citep{2002MNRAS.333..847S}. The monotonic variation of an
initial condition (e.g. $p_z$) may lead either to quasi-periodic orbits around
stable periodic orbits and to the sticky to them chaotic orbits, or to direct
diffusion in the chaotic sea. This happens in general in a nonmonotonic way. As
an example, we give in Fig.~\ref{nonmon} the orbit of the model of snapshot 3
with $(E_J, y_0)=(-0.14,1.4)$, given in Fig.~\ref{sixorb3}e, perturbed by
$p_z=0.038, 0.039$ and $0.040$.
We observe that for $p_z=0.038$ the orbit clearly diffuses in the configuration
space, for $p_z=0.039$ has a double boxy character for more than 3~Gyr before
starting diffusing in the chaotic sea (the quantity $\log_{10}$(GALI$_2$)
$\lessapprox -10$ at $t=10$~Gyr) and for $p_z=0.040$ we have again a practically
chaotic orbit.

\begin{figure*} 
\begin{center}
\includegraphics[scale=0.75]{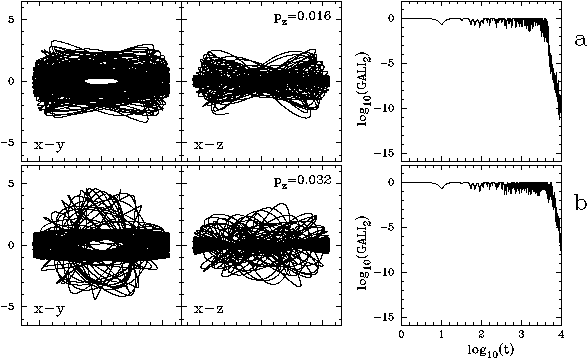}\\
\end{center}
    \caption{The planar orbit presented in Fig.~\ref{fiveo}e ($(E_J,
y_0)=(-0.162,1.27)$) perturbed by $p_z=0.016$ (a) and $p_z=0.032$ (b). The
orbit has initially a x1v3 side-on profile, while for longer times its
morphology is different in each case.} 
    \label{pert4e}
\end{figure*}

In order to demonstrate the fact that at energies where several families
of 3D periodic orbits co-exist, different perturbations of the planar orbits may
lead us to different boxy configurations, we present in Fig.~\ref{pert4e} the
planar orbit of the model of snapshot 2, given in Fig.~\ref{fiveo}e. It has
$(E_J, y_0)=(-0.162,1.27)$. In (a) $p_z=0.016$. The orbit, being initially boxy
on the equatorial plane, has a narrow side-on profile. This morphology lasts for
more than 3~Gyr. Then the orbit occupies a larger volume in phase space.
However, it retains a less confined, but boxy, character in its face-on view,
while in the side-on view its morphology resembles the one encountered in orbits
close to the stable ``frown'' and ``smiles'' periodic orbits. 
For $p_z=0.032$ (Fig.~\ref{pert4e}b) the orbit remains boxy on its face-on view
for even longer time than for $p_z=0.016$, but then diffuses in phase space and
does not have any particular morphology in either projections. The side-on views
of the orbits in both cases of Fig.~\ref{pert4e}, clearly indicate that they
have been trapped close to a x1v3 periodic orbit, which is bifurcated at the 3:1
vertical resonance \citep{2002MNRAS.337..578P}. It is worth to underline that
the side-on profiles of the double boxy orbits in which a morphology of a higher
order $n:1$ resonance may be identified are in general narrower as we approach
corotation, in agreement with the profile of the corresponding periodic orbits
found in \citet{2002MNRAS.337..578P}.

Before closing, we want to add a comment on the general morphology of the three 
models from the MM simulation and especially in the one that appears in 
snapshots 3 and 4 (cf. figure 1 in MM). This morphology is one of a bar 
surrounded by a ring, with the areas on the sides of the bar being rather 
depleted from particles. Beyond this central structure there is a disk without 
any special feature. The bar and ring morphology could easily correspond to that 
of a bar with an inner ring \citep{1996FCPh...17...95B}. However, in this 
particular case corotation is far away, so the question that arises is what is 
the orbital content behind this structure in the model. The orbits that we have 
presented so far support a bar of the size of the bars in the MM $N$-body 
snapshots. The folding of the characteristic provides appropriate round orbits 
with the right dimensions to support the ring. 

\begin{figure} 
\begin{center}
\includegraphics[scale=0.50]{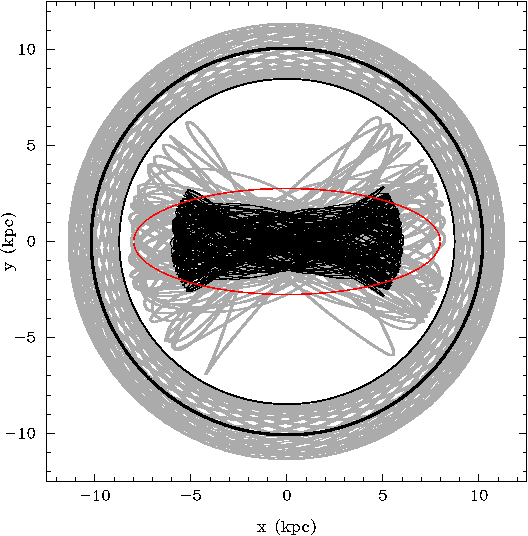}\\
\end{center}
    \caption{The set of the four non-periodic orbits of the model for snapshot 
4 that are depicted in Fig.~\ref{boxring4}
reproduce the basic morphological features of the corresponding MM model. Two 
circular periodic orbits at close-by energies are also plotted. A ringed bar 
morphology is formed, however away from corotation, which in this case is at  
22.89~kpc. The red ellipse indicates the bar in the MM snapshot.}
    \label{mm1}
\end{figure}

Focusing on the model of snapshot 4, we can see that the sticky bar-supporting 
orbits have face-on projections that reinforce a bar structure with two minima 
along the minor axis, giving it a bow-like or peanut-like form, however on the 
equatorial plane of the model in this case (Fig.~\ref{m4threed}a). The planar, 
sticky, boxy orbits have a similar shape (Fig.~\ref{boxring4}a). The 
corresponding MM model has this morphological feature as well. At larger 
energies one can find in the plane some tumbling bar-supporting orbits (see 
Fig.~\ref{boxring4}b). However, even by considering these orbits to be among 
those that populate the model, the areas on the sides of the bar remain rather 
empty. Finally, not only the presence of the circular orbits with the right 
dimensions, but also the regression of the characteristic and the following 
continuation of the curve forwards, i.e. towards corotation, favors the 
accumulation of round orbits around the bar. This folding of the characteristic 
brings in the system twice as much stable circular periodic orbits as in the 
rest of the energies and this supports the formation of a circular ring at a 
certain distance. All these are summarized in Fig.~\ref{mm1}, where we combine 
the orbits of Fig.~\ref{boxring4} in order to reproduce the main morphology of 
the $N$-body MM model. We also plot two circular periodic orbits as a reference 
to the dimensions of the ring. The initial conditions for the two periodic 
orbits are: $E_J=-0.0961538$, $y_0=8.49173$ and $E_J=-0.0898683$, $y_0=10.0141$, 
respectively. Fig.~\ref{mm1} is by no means the result of a 
self-consistent Schwarzschild-type model. It just shows that in the snapshot 
``4'' model, exist orbits that can reproduce the morphology of the corresponding 
MM $N$-body model. 

\vspace{0.5cm}
\subsection{Fast rotating bars}
In the present paper we studied orbital boxiness in the MM models, all
of which have slow rotating bars. We have found that boxiness is a
property associated with the presence of x1 orbits supporting the
bar. This is not related with the pattern speed of the model per
se. However, in slow rotating models like the models of the MM
simulations, there are a lot of non-bar-supporting orbits between the
end of the bar and corotation. These are the circular
orbits. Contrarily, in fast rotating bars one can find bar supporting
elliptical x1 orbits almost all the way from the center of the system
to corotation.

In order to examine the dependence of the results on the pattern speed 
a systematic study with models of bars rotating in a range of $\Omega_b$ is 
needed. This is not done in the present paper. Nevertheless, we considered the 
potential of one of the models (model of snapshot "3") with a higher pattern 
speed, so that we obtained a ratio $R_{CR}/\alpha_b=1.1$. This model is not the 
result of an N-body simulation. It has been used just for studying the orbital 
behavior of bar supporting orbits close to corotation.

We found again in this case that the 2D orbits at the borders of the 
stability islands of x1 were boxy and we could find 3D boxy orbits by perturbing 
them in the vertical directions within a certain $\Delta p_z$ range. So the rule 
in principle applies also in the case of bar supporting orbits close to 
corotation. However, we have to note that the planar boxy bar-supporting orbits 
we could find close to corotation on the plane were like the orbit "1" in 
Fig.~\ref{forbs} and not like orbit "3" of the same figure. In other words at 
their apocentra the segments that were parallel to the minor axis of the bar 
were relatively small.  On the $(y,p_y)$ Poincar\'{e} sections we found more 
islands of orbits of higher multiplicity than in Fig.~\ref{sos2}. Many of them 
surround the stability islands of x1 and this affects the shape of the sticky 
orbits in the region. Also the $\Delta p_z$ range for which we could find double 
boxy orbits was much smaller. The 3D double boxy bar-supporting orbits remained 
confined close to the equatorial plane. We could find orbits with boxy edge-on 
profiles away from the equatorial plane, but their face-on projections did not 
support the bar.

In conclusion: The mechanism applies independently of the pattern
speed value. However it applies more efficiently away from
corotation. If a bar stops away from corotation (as in the slow
rotating models) then almost all of it can be considered as a double
boxy structure. In fast rotating models the 3D double boxy part can be
found pronounced in the inner parts of the bar.

\section{DISCUSSION AND CONCLUSIONS}
\label{sec:conclusions}

The orbital analysis we present in this paper, suggests a recipe for building
two- and three-dimensional boxy structures in rotating bars. The basic
idea is the following: Let us start with the planar backbone of periodic orbits
for building a bar, namely with the well known x1 family. However, instead of
populating the model with regular quasi-periodic orbits encountered in the
immediate neighborhood of the periodic orbit, we consider either periodic orbits
close to the last KAM or, more efficiently, the sticky orbits that surround the
islands of stability, as they appear in the surfaces of section. The
selection of these orbits secure a boxy morphology on the plane. 

In a 3D model, when we eject out of the plane particles that follow the 2D boxy
orbits by adding a $p_z\neq 0$ perturbation, we find that there is always a
$\Delta p_z$ range of perturbations for which all three projections of the 3D
orbits are boxy. A remarkable property of these sticky boxy orbits is the
formation of an X feature embedded in the bar in the \textit{face-on}
projections.

In several cases the side-on views had a peanut-shaped morphology. However, it
is beyond the scope of the present paper to attribute specific orbits, or sets
of orbits to the observed peanut shapes encountered in edge-on galaxies or
snapshots of $N$-body models. This was investigated thoroughly in
\citet{2014MNRAS.445.3546P}. In this study we emphasize that as long as we have
the usual ellipses of the x1 family \citep[or the x1-tree in 3D models according
to][]{2002MNRAS.333..847S} in a rotating bar, we can find a class of
boxy 2D and 3D orbits. They are sticky chaotic orbits as their GALI$_2$ index
indicates and they can support the bar, or a part of the bar, for many Gyr. 

Observational features that can be reproduced by using such orbits as building
blocks, can firstly be the boxy- or peanut-shaped bulges in the central parts of
the bars. In these cases in the face-on views of the galaxies, we will observe
boxy isophotes in their central parts, inside the bar, as in the sample of
galaxies presented by \citet{2013MNRAS.431.3060E}. On the other hand,
the
present study indicates that in cases of slow rotating bars as in the MM
models, the 3D boxy structure may constitute a major part of
the bar. The presence of the X feature in the face-on views of the orbits, as
well as the presence of a ring surrounding the bar, raises the question whether
a dynamical mechanism as the one proposed by \citet{2015MNRAS.448.3081T}, acts
in galaxies like IC~5240 presented in \citet{2007dvag.book.....B}.

Closing, we enumerate below our conclusions:
\begin{enumerate}
 \item In models where the family of x1 ellipses exists in the MM models we can 
find a class of sticky chaotic orbits with a 2D and/or 3D boxy structure. The 
shape of these orbits, after integrating them for 10~Gyr and the evolution of 
their GALI$_2$ index  show that they can be used as building blocks for 
structures that last for several Gyr. They exist in a large range of $E_J$'s.
\item 2D non-periodic boxy orbits can be found on the outermost
invariant curves around x1 on a surface of section, or in regions in the
immediate neighbourhood of the stability islands. We found them in all $E_J$'s
we encounter x1 periodic orbits that do not exceed the size of the $N$-body bar.
\item For finding 3D orbits with boxy morphology in both face-on and
edge-on views, one has to perturb in the vertical direction the boxy planar
orbits. There is always a $\Delta p_z$ interval in the initial conditions of the
perturbed, initially planar, orbits in which the 3D orbits will have a boxy
structure. These are 3D sticky chaotic orbits. Their face-on projections are
different from those of the quasi-periodic orbits close to x1 and its 3D
bifurcations, at all $E_J$'s we find them.
\item In the \textit{face-on} projections of these sticky boxy orbits we find
the formation of an X embedded in the boxy structure. \item Such orbits can be
used to construct models with boxy isophotes inside the face-on views of the
bars. The areas of the boxy isophotes in these cases correspond to the extent of
the edge-on boxy bulges, in agreement with the result of
\citet{2014MNRAS.445.3546P}.
\item According to our analysis, the degree of boxiness of a bar, or of
a part of it, indicates which orbits are populated. If quasi-periodic orbits in
the immediate neighbourhood of the periodic orbits of the central family
prevail, the face-on projections will be elliptical. On the other hand if the
majority of the non-periodic orbits building the bar, or its part, are at the
edges of the stability islands and/or sticky chaotic orbits next to them, then
the supported shape in the face-on views will be boxy. In both cases we can have
boxy edge-on profiles.
\item In the case of \textit{slow rotation}, our 3D sticky, boxy orbits can
build boxy bars (not just boxy features embedded in the bars). In such
cases almost the whole bar is boxy. The slow rotation of the models favours the
appearance of a ringed bar morphology, despite the fact that corotation is at
large distances.

On the other hand, in a fast rotating case we examined, we found that 
boxy \textit{bar-supporting} planar orbits close to corotation had small 
segments parallel to the minor axis of the bar at their apocentra. By 
perturbing them in the vertical direction we could find boxy orbits supporting 
the bar confined close to the equatorial plane. In such a case double boxy 
structures are found mainly embedded in the bar. 
\end{enumerate}

%
%
%
%
%
%
 
\acknowledgments{L.C.V. and I.P. thank the Mexican Foundation CONACYT for grants
that supported this research. L.C.V. thanks the Research Center for Astronomy
(RCAAM) of the Academy of Athens for its hospitality during his visit there,
when part of this work has been completed. This work is part of the project
``Study of stellar orbits and gravitational potentials in galaxies, with
numerical and observational methods'' in which researchers from RCAAM and INAOE
participate. Ch.S. acknowledges support by the National Research Foundation of
South Africa (Incentive Funding for Rated Researchers, IFRR) and also thanks
RCAAM for its hospitality during his visits there in order to collaborate with
P.A.P and L.C.V. We acknowledge fruitful discussions with G.~Contopoulos and
L.~Athanassoula. }

\vspace{1cm} 
\bibliography{paper}

\begin{appendices}

\end{appendices}

\end{document}